\documentclass[english,12pt,aps,prd,a4paper,preprintnumbers,floatfix,nofootinbib,showpacs,superscriptaddress, notitlepage]{revtex4-1} 
 \pdfoutput=1
\usepackage[usenames,dvipsnames]{color}  
\usepackage{graphicx}
\usepackage{pifont}
\usepackage{caption}
\usepackage{subcaption}
\usepackage{amsmath}
\usepackage{amssymb}
\usepackage[colorlinks=true,citecolor=darkred,urlcolor=darkred, pdfborder={0 0 0}]{hyperref}
\usepackage[normalem]{ulem}

\usepackage[T1]{fontenc}
\usepackage[latin9]{inputenc}
\usepackage{array}
\usepackage{booktabs}
\usepackage{mathrsfs}
\usepackage{multirow}
\usepackage{tabularx}
\usepackage{bbold}
\usepackage{mhchem}

\usepackage{hyperref}
\hypersetup{colorlinks=true,pdfborder={0 0 0},allcolors=[rgb]{0,0.0,0.4}}

\newcommand{\hs}{\hspace*{0.5cm}}

\newcommand{\be}{\begin{equation}}
	\newcommand{\ee}{\end{equation}}
\newcommand{\bea}{\begin{eqnarray}}
	\newcommand{\eea}{\end{eqnarray}}
\newcommand{\bit}{\begin{itemize}}
\newcommand{\eit}{\end{itemize}}
\newcommand{\nn}{\nonumber}
\newcommand{\crn}{\nonumber \\}

\newcommand{\bet}{\beta}
\newcommand{\ga}{\gamma}

\newcommand{\om}{\omega}

\newcommand{\fr}{\frac}
\newcommand{\bc}{\begin{center}}
\newcommand{\ec}{\end{center}}

\newcommand{\de}{\delta}
\newcommand{\De}{\Delta}

\newcommand{\si}{\sigma}

\newcommand {\ba}{\begin{array}}
	\newcommand {\ea}{\end{array}}
\newcommand{\ben}{\begin{enumerate}}
	\newcommand{\een}{\end{enumerate}}

\begin{document}

\title{ \boldmath
	COHERENT constraints on $Z'$ in  331$\beta$ model.  }

\author{D. T. Binh}\email{dinhthanhbinh3@duytan.edu.vn}
\affiliation{\small Institute of Theoretical and Applied Research, Duy Tan University, Hanoi 100000, Vietnam}
\affiliation{	Faculty of Natural Science, Duy Tan University, Da Nang 550000, Vietnam\\}
\author{L.~T.~Hue}\email{lethohue@duytan.edu.vn}\affiliation{Institute for Research and Development, Duy Tan University, Da Nang City 50000, Vietnam}
\affiliation{Institute of Physics,  Vietnam Academy of Science and Technology, 10 Dao Tan, Ba
	Dinh, Hanoi 100000, Vietnam}

\author{ V. H. Binh}\email{vhbinh@iop.vast.ac.vn}
\affiliation{Institute of Physics, Vietnam  Academy of Science and Technology, 10 Dao Tan, Ba Dinh, Hanoi, Vietnam}
\affiliation{ Graduate University of Science and Technology,
	Vietnam Academy of Science and Technology,
	18 Hoang Quoc Viet, Cau Giay, Hanoi, Vietnam}

\author{D. V. Soa}\email{dvsoa@hnmu.edu.vn}
\affiliation{Faculty of Natural Sciences and Technology, Hanoi Metropolitan University, 98 Duong Quang Ham, Cau Giay, Hanoi, Vietnam\\}

\author{ H. N. Long}\email{hnlong@iop.vast.ac.vn}
\affiliation{Institute of Physics, Vietnam  Academy of Science and Technology, 10 Dao Tan, Ba Dinh, Hanoi, Vietnam}

\date{\today}

\begin{abstract}
	
	We investigate coherent-elastic neutrino-nucleus scattering ($CE\nu NS$)  in 3-3-1 models for different values of  $\beta$ in which $\beta$ is a parameter used to define the charge operator of the 331 models. We show that the number of events predicted by 331$\beta$ model is in agreement with the data given by COHERENT experiment. We evaluate the sensitivity of the mass of Z' boson with 90\% confidence level (CL) and  find  that $m_{Z'}\geq 1.4 $TeV for $\beta=-\sqrt{3}$ with 90\% C.L . We perform $\chi^2$ fit for liquid Argon, Germanium and NaI detector subsystems, we obtain $m_{Z'} \geq [2,3.1 ]$ TeV  with 90\% CL. Our results  indicate low-energy high-intensity measurements can provide a valuable probe, complementary to high energy collider searches at LHC and electroweak precision measurements.	
\end{abstract}


\maketitle

\section{Introduction}
\label{intro}

Coherent elastic neutrino-nucleus scattering ($CE\nu NS$) is the process where an incident neutrino interacts coherently with the  nuclei.  $CE\nu NS$ was first proposed by Freedman \cite{Freedman1974} about fifty years ago. In  $CE\nu NS$, the interaction of neutrinos and quarks through Z-boson exchange gives a coherent interaction between neutrino and the nucleus as a whole  \cite{Drukier1984} therefore   the cross section is proportional to the quadratic of the number of nucleons A.  The coherent scattering happens when the transferred momentum $q$ is small compared with the atom size, $qR \leq 1$, with $R$ is the nuclear radius. The typical inverse sizes of most nuclei are in the range from 25 to 150 MeV. Hence $CE\nu NS$'s conditions  can be satisfied for reactor neutrinos and play an important role in astrophysical environment like   supernovae and neutron stars  \cite{Freedman1977}.

The COHERENT collaboration  \cite{Akimov2017} observed $CE\nu NS$ for the first time by using a 14.6-kg CsI[Na] scintillating detector with a 4.2 keV energy threshold exposed to the neutrino flux generated at the Spallation Neutron Source (SNS) at Oak Ridge National Laboratory. The  CE$\nu$NS process was observed at a 6.7-$\si$ confidence level (CL),  in agreement with the Standard Model(SM) prediction at  1-$\si$ level. The $CE\nu NS$ data can be used to study other types of physics beyond Standard Model (BSM) such as non-standard neutrino interaction (NSI) \cite{Barranco2005,Barranco2007,Billard2018,Lindner2017,Dent2017,Shoemaker2017,Coloma2017,Flores2020,Miranda2020,Scholberg2006}, sterile neutrino
\cite{Anderson2012,Dutta2016,Kosmas2017}
, neutrino magnetic moment \cite{Dodd1991,Kosmas2015}, light dark matter \cite{2015} or additional neutral gauge bosons \cite{Abdullah2018,Dutta2016a,Miranda2020a,Crivellin:2021bkd}.

A new natural gauge boson $Z'$ will appear naturally in some  gauge extensions   of the SM such as the  Left-Right symmetric model   \cite{Senjanovic1979,Senjanovic1975},  the model of composite boson \cite{Baur1987} , and the 3-3-1 models \cite{Pisano1992,Frampton1992,Foot1993,Singer1980,Foot1994,Montero1993,Long1996,Long1996a,Diaz2005,Diaz2004,CarcamoHernandez2019,Buras:2012jb,Long2019}. They belong to a  class of the  $SU(3)_L$ gauge extensions of the SM,
where the SM  fermion doublets are embedded  in  $SU(3)_L$ triplets or antitriplets including new exotic   fermions  in the third components of the $SU(3)_L$ (anti) triplets. The new exotic fermion in the bottom component leads to the fact that the charge operator is identified by a degree of freedom which is the parameter $\beta$. 

There have been works on the bound for the mass of 
$Z^\prime$ boson in the 3-3-1 models:

\begin{itemize}
	\item[-] The dark matter direct search \cite{Profumo:2013sca} give the lower bound for $Z'$,  $m_{Z'}\geq 2$ TeV.
	
	\item[-] The muon anomalous magnetic moment (g-2) is one of the most precise measurement in physics has been studied in the 3-3-1 models framework.  It is shown that non of 3-3-1 models can address the 4.2$\sigma$ \cite{Muong-2:2021ojo,Aoyama:2020ynm}  discrepancy between SM and experiment data \cite{Ky:2000ku,Kelso:2013zfa,Binh:2015cba,DeConto:2016ith} since the symmetry breaking of $SU(3)_L$ needs to be at scale $\sim 1$ TeV to explain $g-2$. However, in recent work by A. S. de Jesus, et. al \cite{DeJesus:2020yqx}, by introducing inert scalar triplet and vector-like leptons and embed  in 3-3-1 models, the g-2 can be neatly addressed in  3-3-1 models. In the case of neutral heavy leptons 3-3-1 model ($\beta=-1/ \sqrt{3}$), method by  A. S. de Jesus, et. al set lower bound on the mass of Z' boson $m_{Z'} \geq 2$ TeV \cite{DeJesus:2020yqx}.
	
	\item[-] The flavour changing neutral current processes (FCNC) in the 3-3-1 models at tree level are dominated by the exchange of $Z'$ boson. Data from rare decays $B_{s,d}\rightarrow \mu^+\mu^-$ and  $B_d \rightarrow B^\ast (B)\mu^+\mu^- $ imposes a lower bound of mass of Z' $m_{Z'}\geq $1 TeV  \cite{Buras:2013dea,Buras:2014yna}. 
	
	\item[-] In the SM, the atomic parity violation (APV) caused by the neutral gauge boson Z, in BSM APV get additional contribution from $Z'$ boson. Recently, APV data \cite{Bennett:1999pd} of Cesium  $\ce{^{133}_{55}Cs}$ and proton set the low value  of $Z'$ boson mass $m_{Z'} \geq 1.27$ TeV \cite{Long2019}.

	\item[-] The most stringent bounds on the mass of $Z'$ based on LHC bileptons resonance search  imposing the mass of $Z'$ $m_{Z'} \geq 3.7$ TeV    \cite{Nepomuceno2020} and deep learning analysis on LHC data give bound of the mass of $Z'$ $m_{Z'} \geq 4.0$ TeV \cite{Cogollo:2020afo}.
	
\end{itemize}

In this work we focus on the 3-3-1 model with an arbitrary parameter $\beta$  (331$\beta$).  In general, the class of 3-3-1 models have the same characteristics as follows:
1) The anomaly in 3-3-1 model is canceled when all fermion generations are considered, 2) Peccei-Quinn (PQ) symmetry~\cite{Peccei1977} is a
result of gauge invariant in the model 3) As the extension of the gauge group there appears new neutral gauge boson $Z'$, 4) One generation of quark is different from the other two ones,  leading to the appearance of the  tree level Flavor Changing Neutral Current (FCNC) through the mixing $Z-Z'$ \cite{Promberger2007,CarcamoHernandez2006}.

Our paper is organized as follows. In Sec. \ref{models}\ we briefly introduce the 3-3-1 model with arbitrary $\beta$ then we study the neutrino and quarks interactions based on effective Lagrangian of four Fermi interaction for this class of model. In Sec. \ref{sect3}  we consider the setup for the COHERENT experiment and evaluate the event rate for 331$\beta$ model. In Sec.\ref{sec4} we perform the numerical analysis and $\chi^2$ test to study the sensitivity of the mass of the $Z'$ boson with given COHERENT data and future experimental setup. Finally Sec.\ref{section5} is for conclusion.

\section{ The model 331$\beta$ }
\label{models}

The model 331$\beta$ is constructed based on the gauge group $SU(3)_c\times SU(3)_L\times U(1)_X$. One common feature of the class  of $SU(3)_L$ models is that the extension of the gauge group from $SU(2)_L\rightarrow SU(3)_L$ requires new fermions. Normally, $SU(2)_L$  doublets are embedded in the $SU(3)_L$ triplets or antitriplets, while the $SU(2_L)$ singlets  is still $SU(3)_L$ singlets or some of them become the bottom components of the triplets.  New left-handed exotic fermions appear as the third components of the $SU(3)_L$ triplets or antitriplets, while the respective right-handed fermions usually are singlets. The anomaly cancellation requires that the number of fermion triplets equals the number of fermion antitriplets,  leading to the consequence that one quark family must have the same $SU(3)_L$ representation as the three lepton families and different from the remaining quark families. The electric charges of all particles in the 331$\beta$ model are determined by the following charge operator

\be
Q  = T_3 + \bet \,  T_8 + X
\label{eq:charge_Q}
\ee

where  $T_3$, $T_8$ are  the $SU(3)$ generators.
The models are characterized by the parameter  $\beta$ in  the charge operator $Q$.
The lepton representation can be represented as follows \cite{Diaz2005,Diaz2004}:
\bea && L'_{aL}=\left(
\begin{array}{c}
	l'_a \\
	-\nu'_{a} \\
	E'_a \\
\end{array}
\right)_L \sim \left(1,3^*~, -\fr{1}{2}+\fr{\beta}{2\sqrt{3}}\right), \hs a=1,2,3,\crn
&& e'_{aR}\sim   \left(1, 1~, -1\right)  , \hs \nu'_{aR}\sim  \left(1, ~1~, 0\right) ,\hs E'_{aR} \sim   \left(1, ~1~, -\fr{1}{2}+\fr{\sqrt{3}\beta}{2}\right).  \label{lep}
\eea

In particular, the left-handed leptons are assigned to anti-triplets while the right-handed leptons to  singlets.  The model predicts three  exotic leptons $E'^a_{L,R}$ which are much heavier than the ordinary 
ones.  The right-handed neutrinos $\nu'_{aR}$ are needed to generate Dirac mass for active neutrinos.
The prime denotes flavor states to be distinguished with mass eigenstates
being introduced later.
The numbers in the parentheses are to label the representation of $SU(3)_C \otimes SU(3)_L\otimes U(1)_X$ group.

The detail calculation of gauge and Higgs interactions
has been shown in Refs. \cite{Diaz2005,Diaz2004,CarcamoHernandez2006,Buras:2012jb}.
The  covariant derivative is defined as follows
\be D_{\mu}\equiv \partial_{\mu}-i g T^a W^a_{\mu}-i g_X X T^9X_{\mu},  \label{coderivative1}
\ee
where $T^9=\fr{\mathbb{1}}{\sqrt{6}}$, $g$ and $g_X$ are coupling constants corresponding to the two groups $SU(3)_L$ and $U(1)_X$, respectively.
The matrix $W^a T^a$  for a triplet
can be written as
\bea W^a_{\mu}T^a=\fr{1}{2}\left(
\begin{array}{ccc}
	W^3_{\mu}+\fr{1}{\sqrt{3}} W^8_{\mu}& \sqrt{2}W^+_{\mu} &  \sqrt{2}Y^{+A}_{\mu} \\
	\sqrt{2}W^-_{\mu} &  -W^3_{\mu}+\fr{1}{\sqrt{3}} W^8_{\mu} & \sqrt{2}V^{+B}_{\mu} \\
	\sqrt{2}Y^{-A}_{\mu}& \sqrt{2}V^{-B}_{\mu} &-\fr{2}{\sqrt{3}} W^8_{\mu}\\
\end{array}
\right),
\label{wata}\eea
where we have denoted
the charged gauge bosons as
\bea W^{\pm}_{\mu}=\fr{1}{\sqrt{2}}\left( W^1_{\mu}\mp i W^2_{\mu}\right),\crn
Y^{\pm A}_{\mu}=\fr{1}{\sqrt{2}}\left( W^4_{\mu}\mp i W^5_{\mu}\right),\crn
V^{\pm B}_{\mu}=\fr{1}{\sqrt{2}}\left( W^6_{\mu}\mp i W^7_{\mu}\right).
\label{gbos}\eea
From (\ref{eq:charge_Q}), the electric charges of the gauge bosons are
given by
\be
A=\fr{1}{2}+\beta\fr{\sqrt{3}}{2}, \hs
B=-\fr{1}{2}+\beta\fr{\sqrt{3}}{2}\label{charge_AB}.\ee

The scalar sector contains
three scalar triplets as follows
\bea && \chi=\left(
\begin{array}{c}
	\chi^{+A} \\
	\chi^{+B} \\
	\chi^0 \\
\end{array}
\right)\sim \left(1, 3~, \fr{\beta}{\sqrt{3}}\right), \hs
\eta=\left(
\begin{array}{c}
	\eta^0 \\
	\eta^- \\
	\eta^{-A} \\
\end{array}
\right)\sim \left(1, 3~, -\fr{1}{2}-\fr{\beta}{2\sqrt{3}}\right)
\crn
&&
\rho=\left(
\begin{array}{c}
	\rho^+ \\
	\rho^0 \\
	\rho^{-B} \\
\end{array}
\right)\sim \left(1, 3~, \fr{1}{2}-\fr{\beta}{2\sqrt{3}}\right),
\label{higgsc}
\eea
where $A,B$ denote electric charges as
determined in (\ref{charge_AB}).  Only the vacuum expectation values (VEV) of the neutral Higgs components are non zero and defined as follows: $\langle  \chi^0\rangle=\om/\sqrt{2}$, $\langle  \rho^0\rangle=v/\sqrt{2}$,  and $\langle  \eta^0\rangle=u/\sqrt{2}$. 

As usual, the symmetry breaking  happens in two steps:
$SU(3)_L\otimes U(1)_X\xrightarrow{ \om} SU(2)_L\otimes U(1)_Y\xrightarrow{v,u} U(1)_Q$.
Therefore, it is reasonable to assume that $\om \gg  v,u$. There are well-known relations between the gauge couplings of the 331$\beta$ model and the SM, namely
\be
g_2 = g,\hs  \fr{g_X^2}{g^2} = \fr{6s_W^2}{1-(1+\beta^2)s_W^2},,
\label{matching_coupl}
\ee
where $g_2$ and $g_1$ are the  couplings
corresponding to $SU(2)_L$ and $U(1)_Y$ subgroups,  respectively. The weak mixing angle is defined as   $ \sin\theta_W \equiv s_W$,  $\tan\theta_W \equiv t_W =  \fr{g_1}{g_2}$,
and so forth.

The equation in (\ref{matching_coupl}) leads to an interesting constraint of the parameter $\beta$:
\be 
|\beta| \le \sqrt{3} , \hs \beta= \pm \frac{n}{\sqrt{3}}, n=0,1,2,3
\label{eq_beta_constraint}
\ee
With the above VEVs,  the charged gauge boson masses are
\bea
m^2_{Y^{\pm A}} = \fr{g^2}{4}(\om^2+u^{ 2}),\hs
m^2_{V^{\pm B}}=\fr{g^2}{4}(\om^2+v^2),\hs
m^2_W = \fr{g^2}{4}(v^2 + u^{ 2})\, .
\label{masga}\eea

The detail about the lepton sector and the Higgs sector have been given in  \cite{Diaz2005,Diaz2004,CarcamoHernandez2006}  therefore for our purpose of this work we will not present it here.

The Yukawa Lagrangian of quark sector is
\bea
L_{Yuk} &= & \lambda^d_{i,a} {\bar Q}_i \rho d_{a,R}+\lambda^d_{3,a} {\bar Q}_3 \eta^\star d_{a,R} \nn \\
&+&\lambda^u_{i,a} {\bar Q}_i \eta u_{a,R}+\lambda^u_{3,a} {\bar Q}_3 \rho^\star u_{a,R} \nn \\
&+&\lambda^J_{i,j} {\bar Q}_i \chi J_{j,R}+\lambda^J_{3,3} {\bar Q}_3 \chi^\star T_{R} + h.c. \label{Yuk-gen} \eea
where  $i=1, 2, 3$ and $\alpha, \beta=1,2 $ are generation indexes. $J_{1,\,2}=D,\,S$  and T are new exotic quarks. 

One can define the mass eigenstates upon rotation through unitary matrices
\be
\left(\begin{array}{c}
	u^\prime_L  \\
	c^\prime_L   \\
	t^\prime_L  \\
\end{array}\right)=S_u^{-1}\left(\begin{array}{c}
	u_L  \\
	c_L   \\
	t_L  \\
\end{array}\right)\,, \hskip 1 cm
\left(\begin{array}{c}
	d^\prime_L  \\
	s^\prime_L   \\
	b^\prime_L  \\
\end{array}\right)=S_d^{-1}\left(\begin{array}{c}
	d_L  \\
	s_L   \\
	b_L  \\
\end{array}\right)\,, \label{rotation}
\ee
where the rotation matrices are unitary
\be
S_u^\dagger S_u=S_u S_u^\dagger=S_d^\dagger S_d=S_d D_d^\dagger=1
\ee
satisfy
\be 
	V_{CKM}=S_u^\dagger S_d
\ee
with  matrix elements denoted as follows
\be
v_{ij} = (S_d)_{ij}, \qquad u_{ij} = (S_u)_{ij}.
\ee

The neutral currents mediated by $Z$ and $Z'$ bosons relating with neutrinos sector and quark $u$ and $d$ used in our calculation are:
\bea
L_{int}^Z &&= \frac{ i g}{2 c_W} Z^\mu  \Big\{ 
\sum_{\ell=e,\mu,\tau} \left[ {\bar \nu}_{\ell \,L} \gamma_\mu \nu_{\ell L}\right], + \Big[ \Big(1-\frac{4}{3}s_W^2  \Big) \bar{q}_{uL}\gamma_\mu q_{uL} -\frac{4}{3}s_W^2\bar{q}_{uR}\gamma_\mu q_{uR}  \Big]  \nonumber \\
&&+\Big[ \Big(-1+\frac{2}{3}s_W^2  \Big) \bar{q}_{dL}\gamma_\mu q_{dL} +\frac{2}{3}s_W^2\bar{q}_{dR}\gamma_\mu q_{dR}  \Big] \Big\} \,. 
\label{LNCZ} 
\eea

The neutral current mediated by $Z^\prime$ is defined as 
\bea
\label{LNCZp}
&& \,L_{int}^{Z^\prime} =  i\fr{g  Z'^{\mu}}{ 2 \sqrt{3} c_W \sqrt{1-(1+\beta^2) s_W^2}}
\times   \Big\{  \sum_{\ell=e,\mu,\tau}  \left[1-(1+\sqrt{3} \beta) s_W^2 \right] {\bar \nu}_{\ell  } \ga_\mu P_L \nu_{\ell }  \crn
&& + \sum_{i,j=1,2,3} \Big\{ \big[-1+(1+{\beta \over \sqrt{3}})s_W^2\big]\de_{ij}({\bar q}_{u})_i \ga_\mu P_L (q_{u})_j+ 2c_W^2({\bar q}_{u})_i \ga_\mu P_L (q_{u})_j u^*_{3i}u_{3j} \nn \\ 
&& + {4 \over \sqrt{3}}\beta s_W^2 \de_{ij}({\bar q}_{u})_i \ga_\mu P_R (q_{u})_j + \big[-1+(1+{\beta \over \sqrt{3}})s_W^2\big] \de_{ij} ({\bar q}_{d})_i \ga_\mu P_L (q_{d})_j \nn \\
&& + 2c_W^2({\bar q}_{d})_i \ga_\mu P_L (q_{d})_j v^*_{3i}v_{3j}  -{2 \over \sqrt{3}}\beta s_W^2  \de_{ij}({\bar q}_{d})_i \ga_\mu P_R (q_{d})_j \Big\} \Big\},
\eea
where $P_L,P_R=\frac{1 \mp \gamma_5}{2}$ are projection operators; $(q_u)_i,  i=1,2,3$ correspond to $ (u,c,t)$, $q_d, i=1,2,3$ correspond to $(d,s,b)$.

The flavor-changing quark interactions however can be confined to the sector of down quark $q_d$ by choosing $S_u=1$ by alignment in the up type quark sector. In such case $V_L=V_{CKM}$ is  parameterized as \cite{Promberger2007}:
\be
V_L=\left(\begin{array}{ccc}
	\tilde c_{12}\tilde c_{13} & \tilde s_{12}\tilde c_{23} e^{i \de_3}-\tilde c_{12}\tilde s_{13}\tilde s_{23}e^{i(\de_1
		-\de_2)} & \tilde c_{12}\tilde c_{23}\tilde s_{13} e^{i \de_1}+\tilde s_{12}\tilde s_{23}e^{i(\delta_2+\de_3)} \\
	-\tilde c_{13}\tilde s_{12}e^{-i\delta_3} & \tilde c_{12}\tilde c_{23} +\tilde s_{12}\tilde
	\tilde s_{13}\tilde s_{23}e^{i(\delta_1-\de_2-\de_3)} & -\tilde s_{12}\tilde s_{13}\tilde c_{23}e^{i(\de_1 -\delta_3)}
	-\tilde c_{12}\tilde s_{23} e^{i \de_2} \\
	-\tilde s_{13}e^{-i\de_1} & -\tilde c_{13}\tilde s_{23}e^{-i\de_2} & \tilde c_{13}\tilde c_{23}
\end{array}\right) \label{VL-param}
\ee

The interested couplings in  the $331\beta$ model are
\be 
L_{int}^{331\beta}=L_{int}^Z+L_{int}^{Z'}
\label{LNC331beta}
\ee

The common $V-A$ form of 
the interaction of neutral gauge boson $Z,Z'$ with fermions given in Lagrangian (\ref{LNCZ}), (\ref{LNCZp}) are written as:

\be
\mathcal{L}_{Z^i ff}=
\fr{g}{2c_w}\bar{ f}\ga_\mu [g^{Z^i}_V(f)-g^{Z^i}_A(f)\ga _5]fZ_i^{\mu}
\ee
where $Z_i=Z, Z'$ and the $g^{Z^i}_V(f)$, $ g^{Z^i}_A(f)$ are given in Table \ref{ZZpff}.

The common $V-A$ form 
of  the interactions of the  neutral gauge  bosons with  $\nu$ and quarks $u, d$ are
\be
\mathcal{L}_{Z^i ff}=\fr{g}{2c_w}\bar{ f}\ga^\mu [g^{Z^i}_V(f)-g^{Z^i}_A(f)\ga_5]fZ^i_\mu\,,
\ee
where $Z^i=Z, Z'$.  The $g^{Z^i}_V(f)$, $ g^{Z^i}_A(f)$ are given in Table \ref{ZZpff}.
\begin{table}[ht!]
\centering
\begin{tabular}{|c|c|c|c|c|}
	\hline
	$f $ & $g^Z_V(f)$ & $g^Z_A(f)$ &$g^{Z'}_V(f)$ &  $g^{Z'}_A(f)$\\
	\hline
	$\nu$ &  $\fr 1 2$ & $\fr 1 2$ &  $  \frac{f(\beta)}{2\sqrt{3}} [1-(1+\sqrt{3}\beta)s_w^2]$ & $  \frac{f(\beta)}{2\sqrt{3}} [1-(1+\sqrt{3}\beta)s_w^2]$\\
	\hline
	$u$ & $\fr 1 2 -\frac{4}{3} s^2_W$ & $ \fr 1 2 $& $  \frac{f(\beta)}{2\sqrt{3}} \Big( [-1+(1+\frac{5}{\sqrt{3}}\beta)s_w^2]   \Big)$ & $  \frac{f(\beta)}{2\sqrt{3}} \Big(  [-1+(1-\sqrt{3}\beta)s_w^2] \Big)$\\
	\hline
	$d$ & $-\fr 1 2 + \frac{2}{3} s^2_W$ & $-\fr 1 2 $& $    \frac{f(\beta)}{2\sqrt{3}}\Big([-1+(1-\frac{ \beta}{\sqrt{3}})s_w^2]+2c_W^2|\tilde s_{13}|^2 \Big)$ & $  \frac{f(\beta)}{2\sqrt{3}} \Big( [-1+(1+\sqrt{3}\beta)s_w^2]  + 2c_W^2|\tilde s_{13}|^2 \Big) $\\
	\hline
\end{tabular}
\caption{The couplings of $Z$ and $Z'$ with  $\nu$ and quarks $u , d $ in the 331$\beta$ model}
\label{ZZpff}
\end{table}	
Here $f(\beta)=\frac{1}{\sqrt{1-(1+\beta^2)s_w^2}}$.

In the low energy limit ($energy \ll  m_Z$) we have four fermion interactions at which interaction of neutrinos and quarks can be described by the effective Lagrangian. The 
expressions  in (\ref{LNCZ}) and (\ref{LNCZp}) can be rewritten as 
\bea
\mathcal{L}_{eff}^Z&=&\sum_{q=u,d} \sqrt{2}G_F  \bar{\nu}\ga^\mu[g_V^Z(\nu)-g_A^Z(\nu)\ga_5]\nu 
.   \bar{q}\ga_\mu[g_V^Z(q)-g_A^Z(q)\ga_5]q \crn
&=&\frac{G_F}{\sqrt{2}}\bar{\nu}\ga^\mu[1-\ga_5]\nu 
.   \bar{q}\ga_\mu[g_V^Z(q)-g_A^Z(q)\ga_5]q= \frac{G_F}{\sqrt{2}}J^\mu_{NC}J_{NC\mu}  
\label{LeffSM}
\eea
and
\begin{align}
\mathcal{L}_{\mathrm{eff}}^{Z'}&=\sum_{q=u,d} \sqrt{2}G_F  \frac{m^2_Z}{m_{Z'}^2}  \bar{\nu}\ga^\mu[g_V^{ Z'}(\nu)-g_A^{Z'}(\nu)\ga_5]\nu 
.  \bar{q}\ga_\mu[g_V^{Z'}(q)-g_A^{Z'}(q)\ga_5]q  \nn \\
&=\sum_{q=u,d} \frac{G_F}{\sqrt{2}}  \frac{m^2_Z}{m_{Z'}^2}\frac{f(\beta)}{\sqrt{3}}[ 1-(1+\sqrt{3}\beta)s_W^2]  \times \bar{\nu}\ga^\mu[1 -\ga_5]\nu 
\times   \bar{q}\ga_\mu[g_V^{Z'}(q)-g_A^{Z'}(q)\ga_5]q.
\end{align}

In the  $331\beta$ framework, the effective Lagrangian relating with the CE$\nu$NS  is:
\be 
\mathcal{L}_{\mathrm{eff}}^{331\beta}= \mathcal{L}_{\mathrm{eff}}^Z + \mathcal{L}_{\mathrm{eff}}^{Z'}=\sum_{q=u,d} \fr{G_F}{\sqrt{2}}    \bar{\nu}\ga^\mu(1 -\ga_5)\nu 
\times    \bar{q}\ga_\mu[G_V(q)- G_A(q) \ga_5]q,
\label{Leff331beta}
\ee 
where we have introduced two new effective $V-A$ couplings $G_{V,A}(q)$ of quarks $q=u,v$:
\be
G_X(q)=g_X^Z(q)+\frac{m_Z^2}{m_{Z'}^2}\frac{f(\beta)}{\sqrt{3}}[1-(1+\sqrt{3}\beta)s_W^2] \times g_X^{Z'}(q) , \; X=V,A. \label{GVq}
\ee
As a consequence, the effective V-A couplings of the proton $p$ and neutron $n$ contributing to the CE$\nu$NS processes are  
\bea
G_X(p)&=&2G_X(u)+G_X(d),    \label{GVp}\\
G_X(n)&=&G_X(u)+2 G_X(d).   \label{GVn}
\eea

\section{Coherent Elastic Neutrino-Nucleus Scattering}
\label{sect3}

The SM prediction for the differential cross section of $CE \nu NS$ for neutrino with energy $E_\nu$ scatter off a nuclear target $(A,Z)$ with recoil energy $E_R$ and ignoring $\Big( \frac{E_R}{E_\nu}\Big)$ term is given as: 
\cite{Freedman1974,Barranco2005,Barranco2007,Billard2018,Lindner2017,Dent2017}
\be 
\frac{d \sigma_{SM}}{dE_R}=\frac{G_F^2}{4 \pi}m_N\Big[(\mathcal{Q}^V_{W})^2  \left( 1-\frac{m_NE_R}{2E_\nu^2}  \right)+(\mathcal{Q}^A_{W})^2  \left( 1+\frac{m_NE_R}{2E_\nu^2}  \right)\Big]F^2(2m_N E_R)
\label{crosssectionSM-1}
\ee 
where $m_N$ is the nuclear mass and  $F^2(2m_NE_R)$ is the nuclear Helm form factor 
 given 
in  \cite{Barranco2007,Kerman2016,Hoferichter:2020osn} as
\be 
F(q^2)=\frac{3}{qR_0} J_1(qR_0)e^{-\frac{1}{2}q^2 s^2}
\ee
where  $J_1(x)$ is the first order spherical Bessel function. $R_0^2=R^2 - 5s^2$, $s=0.5 fm$ and $R=1.2 A^{1/3} fm$.

The vector  and axial vector  weak charge $\mathcal{Q}^V_{W}, \mathcal{Q}^A_{W}$   are defined as \cite{Barranco2005,Barranco2007,Billard2018,Lindner2017,Dent2017}:
\bea 
\mathcal{Q}^V_{W}&=&-2[Zg^Z_V(p)+Ng^Z_V(n)]=[N-(1-4s_W^2)Z] \\
\mathcal{Q}^A_{W}&=&-2[g^Z_A(p)(Z_+-Z_-)+g^Z_A(n)(N_+-N_-)]
\label{SMcharge}
\eea
where $Z_{\pm}, N_\pm$ denote the number of protons and neutrons with spin up(+) and spin down (-) respectively. For most nuclei, the
 ration $\frac{\mathcal{Q}^A_{W}}{\mathcal{Q}^V_{W}} \approx \frac{1}{A}$ while for spin zero nuclei $\mathcal{Q}^A_{W}=0$. Hence the contribution of axial vector weak charge is  ignored in this work. It means that from now on we will use the following notations in the SM limit:
\be \label{eq_QW}
\mathcal{Q}^V_{W}\equiv 	\mathcal{Q}^{\mathrm{SM}}_{W}, \; \mathcal{Q}^A_{W}=0. 
\ee 
The differential cross section of   CE$\nu$NS predicted by the SM is then given as: 
\be 
\frac{d \si^{\mathrm{SM}}}{dE_R}=\fr{G_F^2}{4\pi}m_N F^2(E_R)\Big[(\mathcal{Q}^\mathrm{{SM}}_{W})^2  \left( 1-\frac{m_NE_R}{2E_\nu^2}  \right)\Big]. 
\label{crosssectionSM-2}
\ee 
This quantity will be used to compare with that predicted by the 331$\beta$. 
\subsection*{CE$\nu$NS from neutrino magnetic moment}

In BSM 
 predicting massive neutrinos, they may have nontrivial interaction with photon through magnetic dipole. In minimal extension of the SM,
   a massive Dirac neutrino may acquire a diagonal  magnetic moment with a magnitude \cite{Fujikawa1980}:
\be 
\mu_\nu \approx3.2\times 10^{-19}\left[\fr{m_\nu}{1\mathrm{eV}} \right]
 \mu_B\,.
\ee 
The masses of 
  neutrinos in  3-3-1 models  
 have been studied 
\cite{Catano2012,Huitu2020,CarcamoHernandez2015,Tully2001}. The GEMMA experiment \cite{Beda2013} 
measuring the $\bar{\nu}_e- e$ scattering has put the strongest constraints on the dipole moment of the reactor neutrino $\bar{\nu}_e$. The limit is $\mu_\nu < 2.9  \times 10^{-11}\mu_B$ (90\% CL). The most stringent astrophysical constraint on $\mu_\nu$ has been recently obtained in \cite{ArceoDiaz2015,Canas2016}  $\mu_\nu< 2.2 \times 10^{-12}\mu_B$.

The cross section for nuclear scattering from the neutrino magnetic moment $\mu_\nu$ is given by \cite{Vogel1989}
as: 
\be 
\fr{d\sigma^{\mathrm{mag}}_{\nu-N}}{dE_R}=\frac{\pi \alpha^2 \mu_\nu^2 Z^2}{m_e^2} \Big( \frac{1}{E_R} -\fr{1}{E_\nu} +\fr{E_R}{4E^2_\nu} \Big)\,.
\ee 

This is the charge-dipole interaction which does not interfere with the CE$\nu$NS by neutral current and  receives a coherence enhancement from the charge of the nucleus and proportional to $Z^2$.

\subsection*{CE$\nu$NS in the 331$\beta$ model}

Since the Lagrangians in (\ref{LeffSM}) and in (\ref{Leff331beta}) have the same structure then in our calculation for the CE$\nu$NS cross section predicted by the model 331$\beta$, it is sufficient to substitute the vector weak charges $\mathcal{Q}^{\mathrm{SM}}_W$ by 
$ \mathcal{Q}^{331\beta}_W$,  where 
\be 
\mathcal{Q}^{331\beta}_W= -2\Big[ZG_V(p)+NG_V(n)\Big],
\label{eq_QW331}
\ee 
$G_V(p)$ and $G_V(n)$ are given in Eqs. in (\ref{GVp}) and (\ref{GVn}), respectively. The  differential cross section is
\be 
\fr{d \sigma^{331\beta}}{dE_R}=\fr{G_F^2}{4\pi}m_N F^2(E_R)\Big[(\mathcal{Q}^{331\beta}_W)^2  \left( 1-\fr{m_NE_R}{2E_\nu^2}  \right)\Big].
\label{crosssectionSM-331}
\ee 
Therefore, the total differential cross section including the parts from neutrino magnetic moment $\mu_\nu$ is 
\be 
\label{eq_sigmatotal}
\fr{d\sigma^{\mathrm{total}}}{dE_R}=  \fr{d\sigma^{331\beta}}{dE_R}  + \fr{d\si^{mag}}{dE_R}. 
\ee

\subsection*{Neutrinos at the Spallation Neutron Source}

The neutrino fluxes coming from the SNS   used by the COHERENT collaboration consist of $\nu_e, \nu_\mu$ and $\bar{\nu}_\mu$. These neutrino are produced by the decay at rest of  $\pi^+ \rightarrow \mu^+ \nu_\mu $ with energy distribution described by \cite{Louis2009,Denton2021}.
\be 
f_{\nu_\mu}= \de \left( E_\nu-\fr{m^2_\pi  -m^2_\mu}{2m_\pi} \right) \,.
\ee 
The $\mu^+$ then decays to antimuon neutrino and electron neutrinos. These neutrinos can be modeled for energies upto 52.8 MeV \cite{Cadeddu2018}.
\bea 
f_{\bar{\nu}_\mu}&&= \fr{64E_\nu^2}{m^3_\mu}\left( \fr{3}{4}-\fr{E_\nu}{m_\mu} \right)\,,  \\
f_{\nu_e}&&= \fr{192E_\nu^2}{m^3_\mu}\left( \fr 1 2 -\fr{E_\nu}{m_\mu} \right) \,.
\eea

The expected number of  $CE\nu NS$ events is given as \cite{Nepomuceno2020}
\be 
\fr{dN}{dE_R}=\frac{N_{target}N_{POT}f_{\nu/p}}{4\pi l^2}\int dE_\nu f_i(E_\nu) \fr{d\si^{Total}_i}{dE_R}(E_\nu) \,.
\ee 
Here $N_{target}$ is the number of the target nuclei. $N_{POT}=1.76 \times 10^{23}$ is the number of protons on target which were
 accumulated during 308.1 days of running time, $l=19.3 m$ is the distance between the detector and the source. $f_{\nu/p}=0.08$ is the  rate of  neutrinos per protons in  collision at SNS. The atomic numbers of the Cs and I nucleus are similar $(A_I = 127, Z_I = 53 ;  A_{Cs} =
133, Z_{Cs} = 55)$. We calculate the individual cross sections separately and weight
the number of events according to the nuclear masses. On TABLE \ref{detector-subsystems} we summarize some experiments with different detectors, threshold energy, efficiency, exposure time and baseline length.

\begin{table}[htbp]
\centering
\begin{tabular}{|c|c|c|c|c|c|c|}
	\hline
	Nuclear& Technology & Mass & Distance from & exposure & effiency &Recoil \\

	target & & (kg) &  source (m) & (day) & &threshold (keVnr)
	\\ \hline 
	
	CsI[Na] & Scintillating crystal & 14.6 & 19.3 &308.1  &\cite{Akimov2018}  & 6.5 \\ 
	\hline
	Ge & HPGe PPC & 10 & 22 & 365& 50\%  & 5 \\
	\hline
	LAr & Single-phase & 24 & 27.5 & 365& \cite{Akimov2019,COHERENT:2018gft, COHERENT:2020iec} &20 \\
	\hline
	NaI[Tl] & Scintillating crystal & 185$^*$/2000 & 28 & 365 &50\% & 13 \\
	\hline
\end{tabular}
\caption{\label{detector-subsystems}Parameters for the  COHERENT
	detector subsystems.}
\end{table}

\section{Numerical results}
\label{sec4}

\begin{figure}[ht!]
\centering
\includegraphics[scale=0.65]{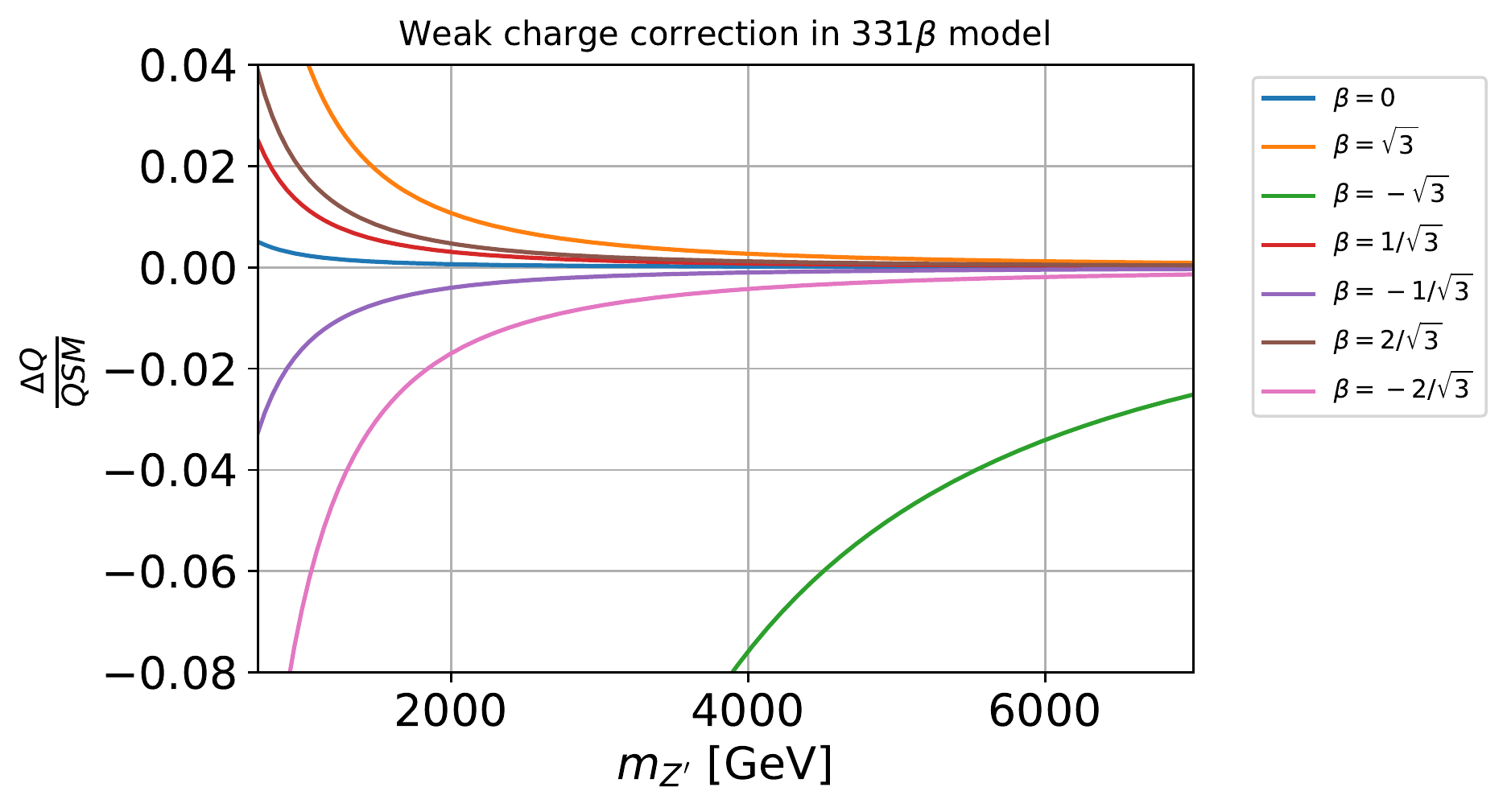}
\caption{The weak charge correction of the 331$\beta$ as a function of $m_{Z'}$ and different $\beta$.}
\label{EQCorr1}
\end{figure}

\begin{figure}[ht!]
\centering
\includegraphics[scale=0.65]{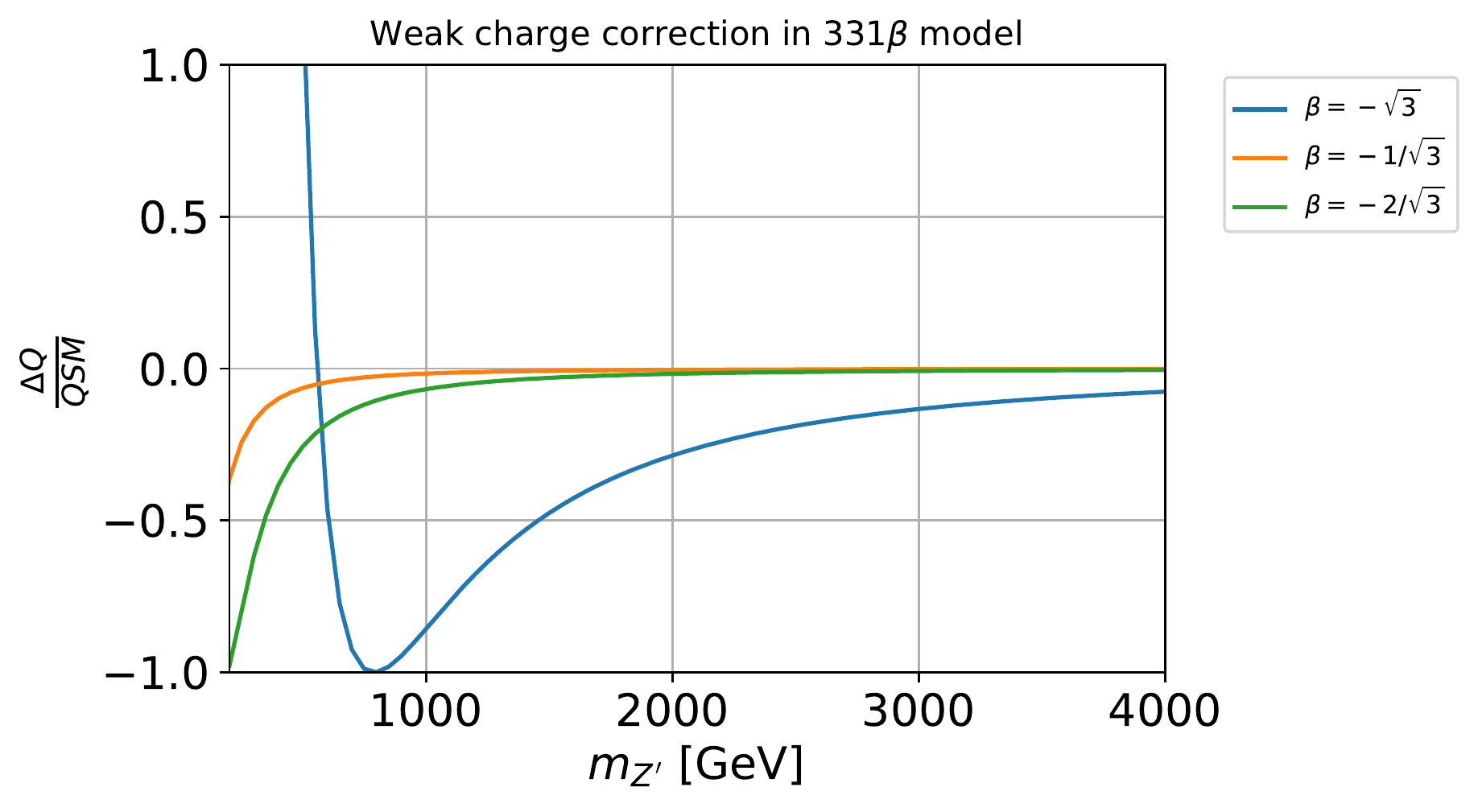}
\caption{The weak charge correction of the 331$\beta$ as a function of $m_{Z'}\le4$ TeV and different $\beta$.}
\label{EQCorr2}
\end{figure}

In the 331$\beta$ model  framework, the parameters space is $\mathcal{P}= \{\beta,  m_{Z'}, \mu_\nu, \tilde{s}_{13} \}$. Since the effective Lagrangians  of the SM  and 331$\beta$ model have the same structure hence the two different cross sections will be scaled by factors of the weak charges $\mathcal{Q}^{\mathrm{SM}}_W$ and $\mathcal{Q}^{\mathrm{331\beta}}_W$. It is convenient to compare the weak charges  of the two models  by defining  the weak charge correction as the following ratio  $\mathcal{R}^{331\beta}_W=   \frac{\De \mathcal{Q}_W^{331\beta}}{\mathcal{Q}_W^{\mathrm{SM}}}=\frac{\mathcal{Q}_W^{331\beta}-\mathcal{Q}_W^{\mathrm{SM}}}{\mathcal{Q}_W^{\mathrm{SM}}}$.  The dependence of this ratio  on the   $Z'$ boson mass  for different values of $\beta$ is shown in Figs.\ref{EQCorr1}
and  \ref{EQCorr2}.
Fig. 1 corresponds to the large range of $m_{Z'}\le 10$ TeV, where we can see clearly that the correction $R^{331\beta}_W$ always has the same sign with $\beta$. In addition, the $|\mathcal{R}^{331\beta}_W|$ will increase with increasing values of $|\beta|=\frac{n}{\sqrt{3}}, n=0,1,2,3 $. In contrast, $|\mathcal{R}^{331\beta}_W|$ decreases with larger $m_{Z'}$ and  approaches zero when $m_{Z'}$ is large enough. This implies a consistent result that   the weak charge  predicted  the 331$\beta$ model approaches the SM value with heavy $m_{Z'}$.   Because the CE$\nu$NS  data is consistent with the SM prediction at the 1-sigma level \cite{Akimov2017},  $m_{Z'}$ must be bounded from below, especially for $\beta=\pm \sqrt{3}$, as we will discuss in detail below.

One of interesting characteristics of the 331$\beta$ model  is that it predicts  the appearance of the tree level FCNC. The weak charge in 331$\beta$ model is therefore  depends on the  mixing parameter $\tilde{s}_{13}$. Using the  data on the mass difference $\Delta M_{d}$ and CP asymmetry   $S_{\psi K_s}$ in the $B_s$ system, 
it was concerned that $\tilde{s}_{13} \leq 0.03$ at $m_{Z'}=3$ TeV \cite{Buras:2012jb}. As a result, the mixing term relating with $\tilde{s}_{13}$ is  of one order smaller than the remaining  part of  the  weak charge defined in Eq.\eqref{eq_QW331}, see more precisely in  table \ref{ZZpff}. In the following analysis we will fix $\tilde{s}_{13} = 0.03$ and focus on the  two specific cases  $\beta=\pm \sqrt{3}$ that give significant deviation on the weak charge and the differential cross section of the CE$\nu$NS.

\begin{figure}[ht!]
\centering
\includegraphics[scale=0.65]{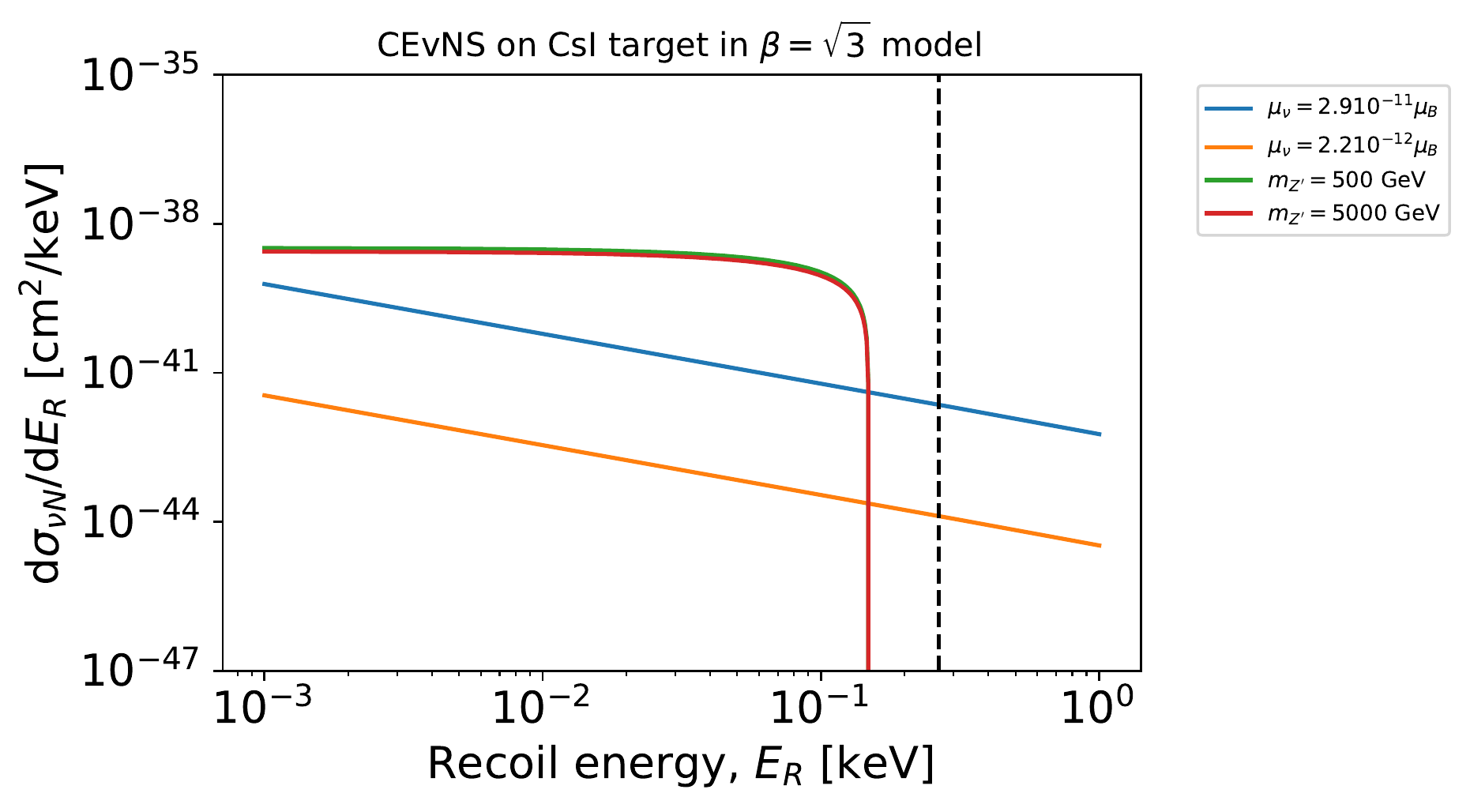}
\includegraphics[scale=0.65]{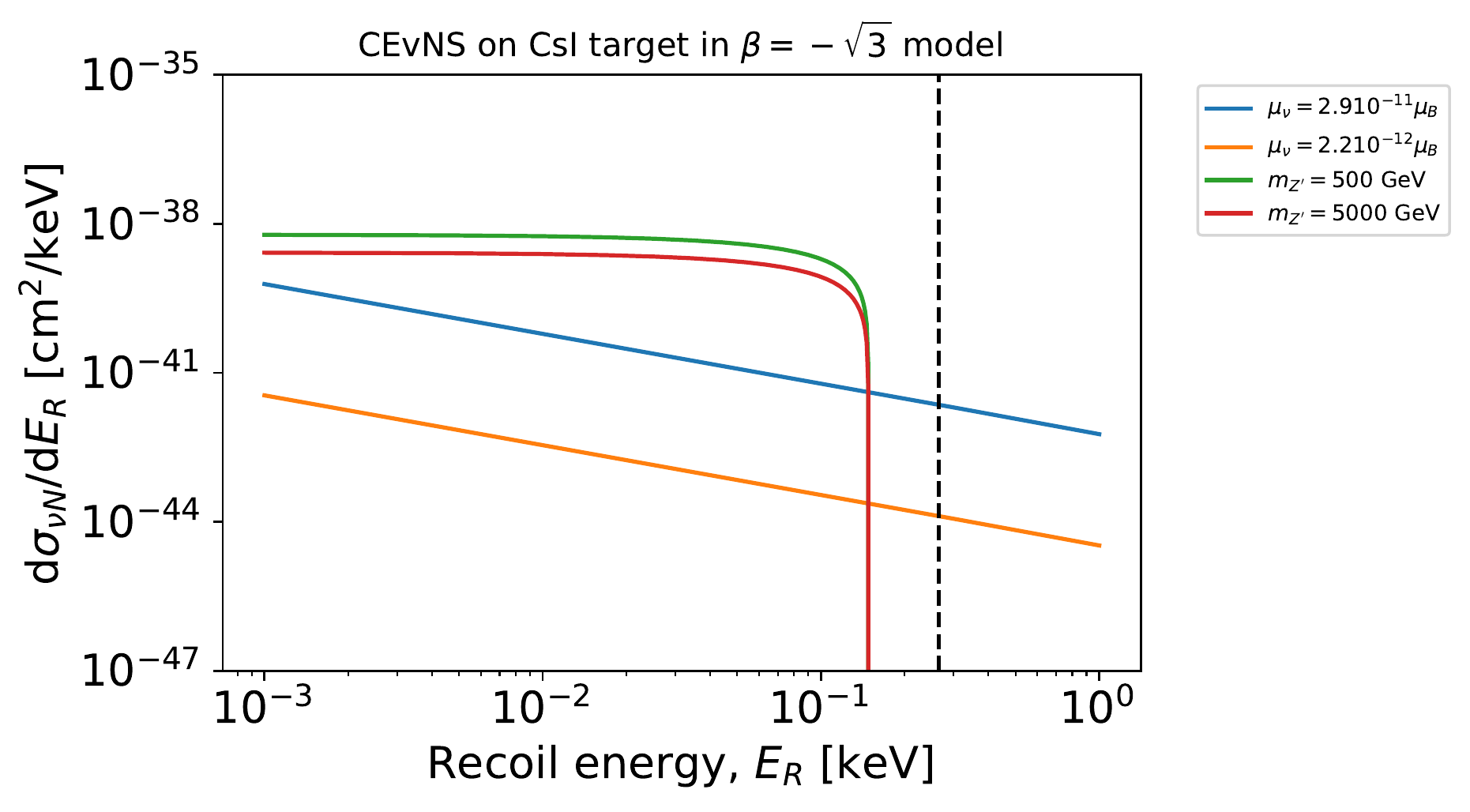}
\caption{The different contributions to the total differential cross sections vs recoil energy}
\label{Diff-Cross-section}
\end{figure}

In the 331$\beta$ model frameworks with active neutrinos having  non-zero Dirac masses, the total CE$\nu$NS cross section at low scattering energies  get contributions from both  parts of neutral current $\frac{d\sigma^{331\beta}}{dE_R}$ and magnetic interaction through dipole moment $\frac{d\sigma^{\mathrm{mag}}}{dE_R}$, as given in Eq.~\eqref{eq_sigmatotal}. These two parts   are shown numerically as functions of $E_R$ in Fig.\ref{Diff-Cross-section}, with $\beta=\pm \sqrt{3}, \tilde{s}_{13}=0.03$ and different $m_{Z'}$ for $\frac{d\sigma^{331\beta}}{dE_R}$; and  $\mu_\nu= \{ 2.9\times  10^{-11}\mu_B, 2.2\times  10^{-12}\mu_B \}$ for $\frac{d\sigma^{\mathrm{mag}}}{dE_R}$. There is an interesting result that the contribution from  neutral current to the total  differential cross section is of orders greater than that from magnetic interaction through dipole moment with  $\mu_\nu=2.9\times  10^{-11}\mu_B $. 

\begin{figure}[ht!]
\centering
\includegraphics[scale=0.5]{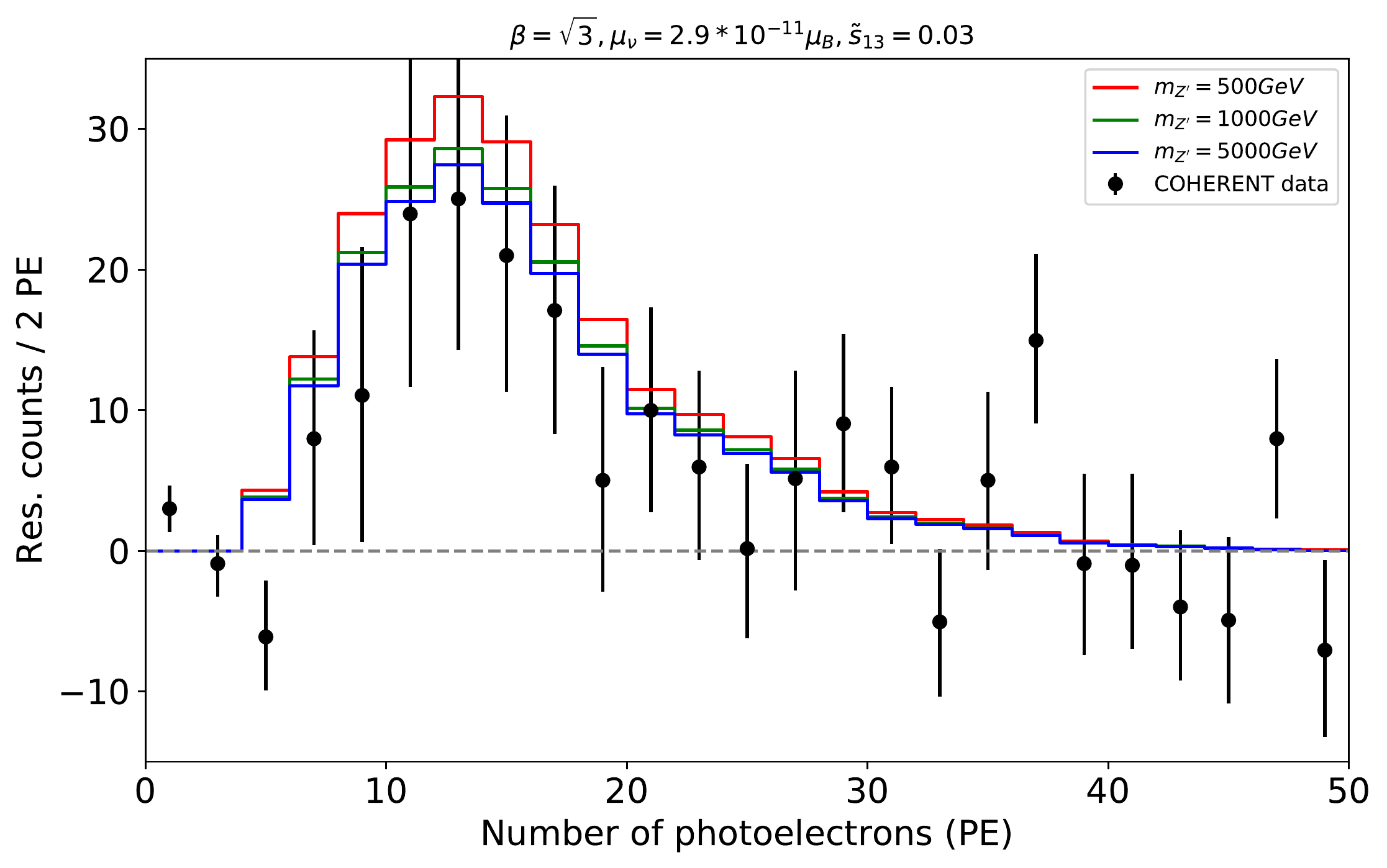}
\includegraphics[scale=0.5]{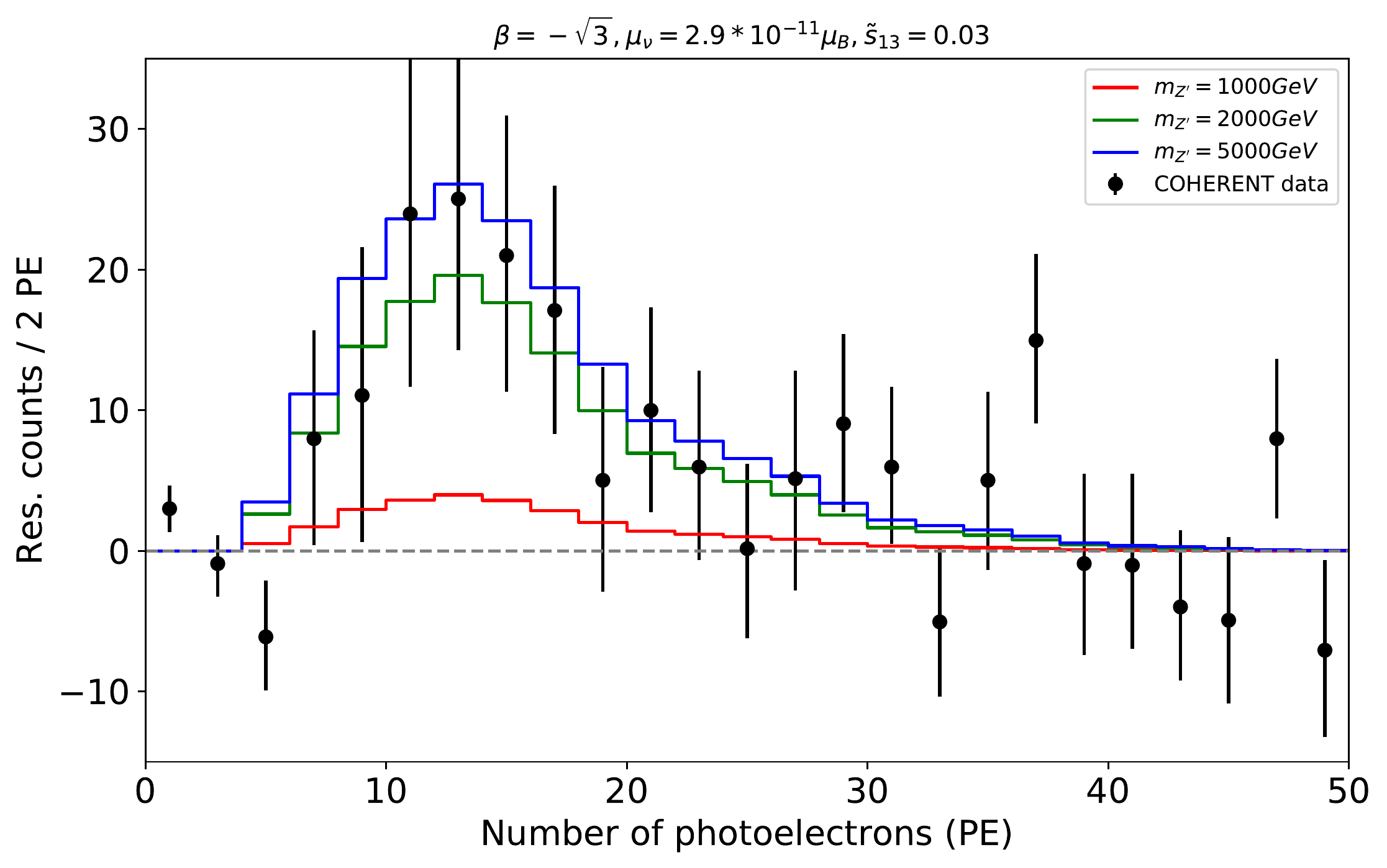}
\caption{Number of events of 331$\beta$ model vs COHERENT data}
\label{331-Coherent Data}
\end{figure}

Next, we calculate the number of events  for different values of $Z'$ boson mass,  $\beta=\pm \sqrt{3}, \;  \mu_\nu=2.9 \times 10^{-11}\mu_B, \; \tilde{s}_{13}=0.03 $, see the numerical results in Table \ref{Nevents}. 
\begin{table}[ht!]

\begin{tabular}{|c|c|c|c|c|c|c|}
	\hline
	&$m_{Z'}$(GeV) & $500$ & $1000$ &$2000$ &  $4000$ &5000 \\
	\hline
	$\beta=\sqrt{3}$ & N events &  222 & 196 &  190 & 189& 188\\
	\hline
	$\beta=-\sqrt{3}$ & N events &  411 & 27 &  134 & 181& 179\\
	\hline
\end{tabular}
\centering
\caption{Number of events for different values of  $m_{Z'}$ with $\beta=\pm \sqrt{3}$ }
\label{Nevents}
\end{table}	
The comparison of these with those in the data given by COHERENT experiment is  shown  Fig.\ref{331-Coherent Data}, where the consistence is found. 

\subsection*{COHERENT constraints on $Z'$ boson mass }
\subsubsection*{CsI detector}

In order to extract the constraints on $Z'$ mass and $\mu_\nu$ from the first phase of COHERENT (with a CsI detector), we compute $ \Delta \chi^2(\mathcal{P})=\chi^2(\mathcal{P})-\chi^2_{min}(\mathcal{P}) $ with $\mathcal{P}= \{ \beta, m_{Z'},\mu_\nu , \tilde{s}_{13}\} $ and  $\chi^2$ is defined in Ref. \cite{Akimov2017},
\be 
\chi^2(\mathcal{P})= \min_{a_1,a_2} \Big[  \fr{(N_{exp}-N_{331\beta}[1+a_1]- B_{on}[1+a_2])^2}{ \si_{stat}^2}+
\left( \fr{a_1}{\si_{a_1}} \right)^2  +\left( \fr{a_2}{\si_{a_2}} \right)^2 \Big],
\ee 
where 
\bit 
\item[-] $N_{331\beta}$ is the number of events predicted by the 331$\beta$ model.

\item[-] $N_{exp}=134$ is the  observed number of  events  in the current COHERENT limits. 

\item[-]  $\si_{stat}=\sqrt{N_{exp}+2B_{ss}+B_{on}}$ is the statistical uncertainty.
\item[-] $B_{on}=6$ is the estimated beam on background.
\item[-] $B_{ss}=405$ is the estimated steady state background. 
\item[-] $a_1$ is the systematic parameter corresponding to uncertainty on the signal rate. $\si_{a_1}$ is the fractional uncertainty corresponding to a 1-sigma variation and is estimated to be $\sigma_{a_1}=0.28$.
\item[-] $a_2$ is the systematic parameter corresponding to uncertainty on the estimate of $B_{on}.\si_{a_2}$ is the fractional uncertainty corresponding to a 1-sigma variation and is estimated to be $\si_{a_1}=0.25$.
\eit

To calculate the $\De\chi^2$ we will first calculate  expected number of events for a given set of parameters $\mathcal{P}= \{\pm \sqrt{3}, 3\;\mathrm{TeV}, 2.9 \times 10^{-11}\mu_B, 0.03 \}$ then minimize  $\chi^2_{min}(\mathcal{P})$ we can obtain $a_1, a_2$. 
The $\De \chi^2(m_{Z'})$ profile is calculated as $ \De \chi^2(\mathcal{P})=\chi^2(\mathcal{P})-\chi^2_{min}(\mathcal{P}) $  and is shown  as in Fig. \ref{331-Coherent Data}. We find that 
the value of $m_{Z'} \geq$ 400 GeV for $\beta=\sqrt{3}$ and $m_{Z'}\geq 1.5$ TeV for $\beta=-\sqrt{3}$ with 90\% CL.

\begin{figure}[ht!]
\centering
\includegraphics[scale=0.55]{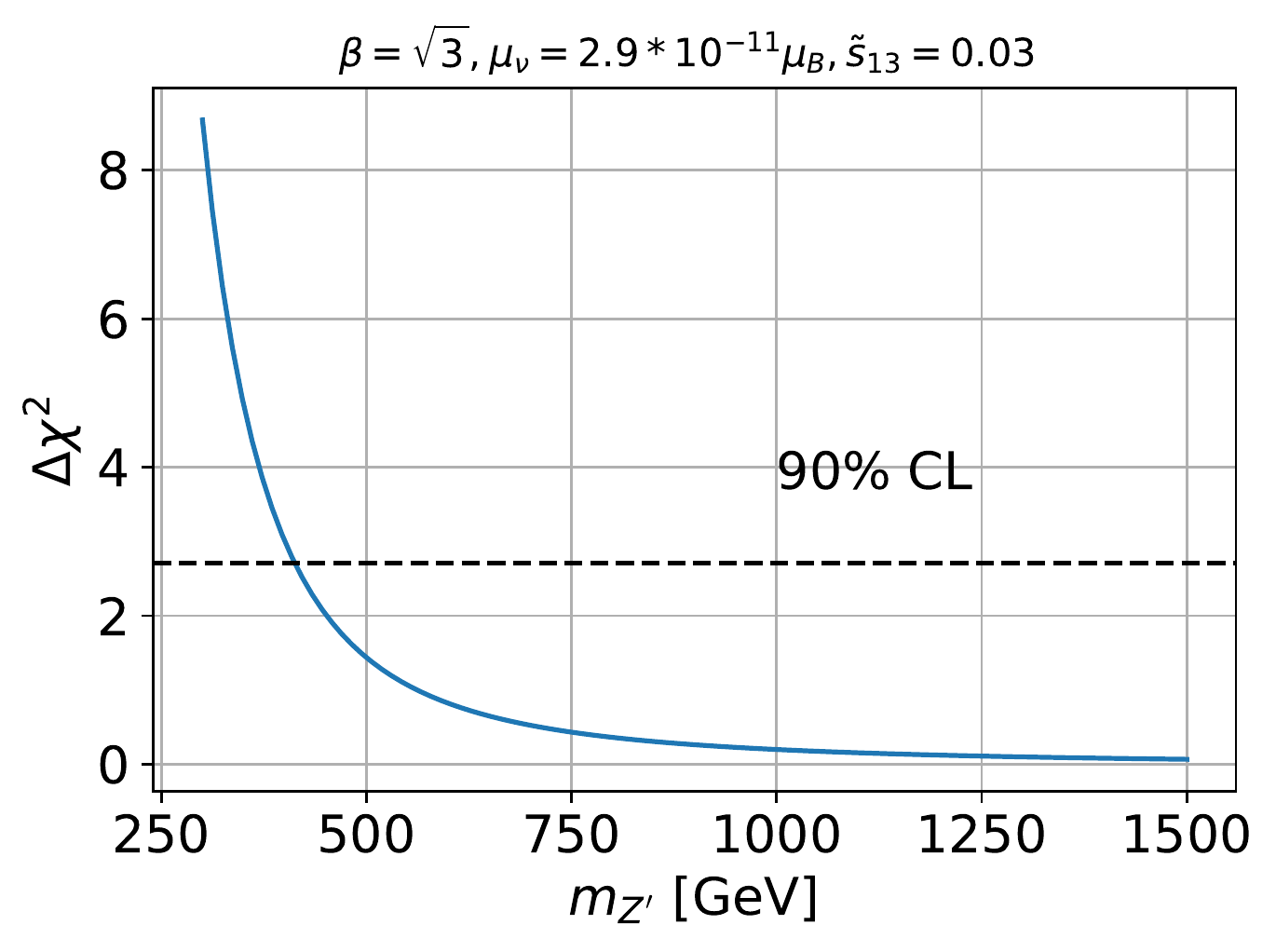}
\includegraphics[scale=0.55]{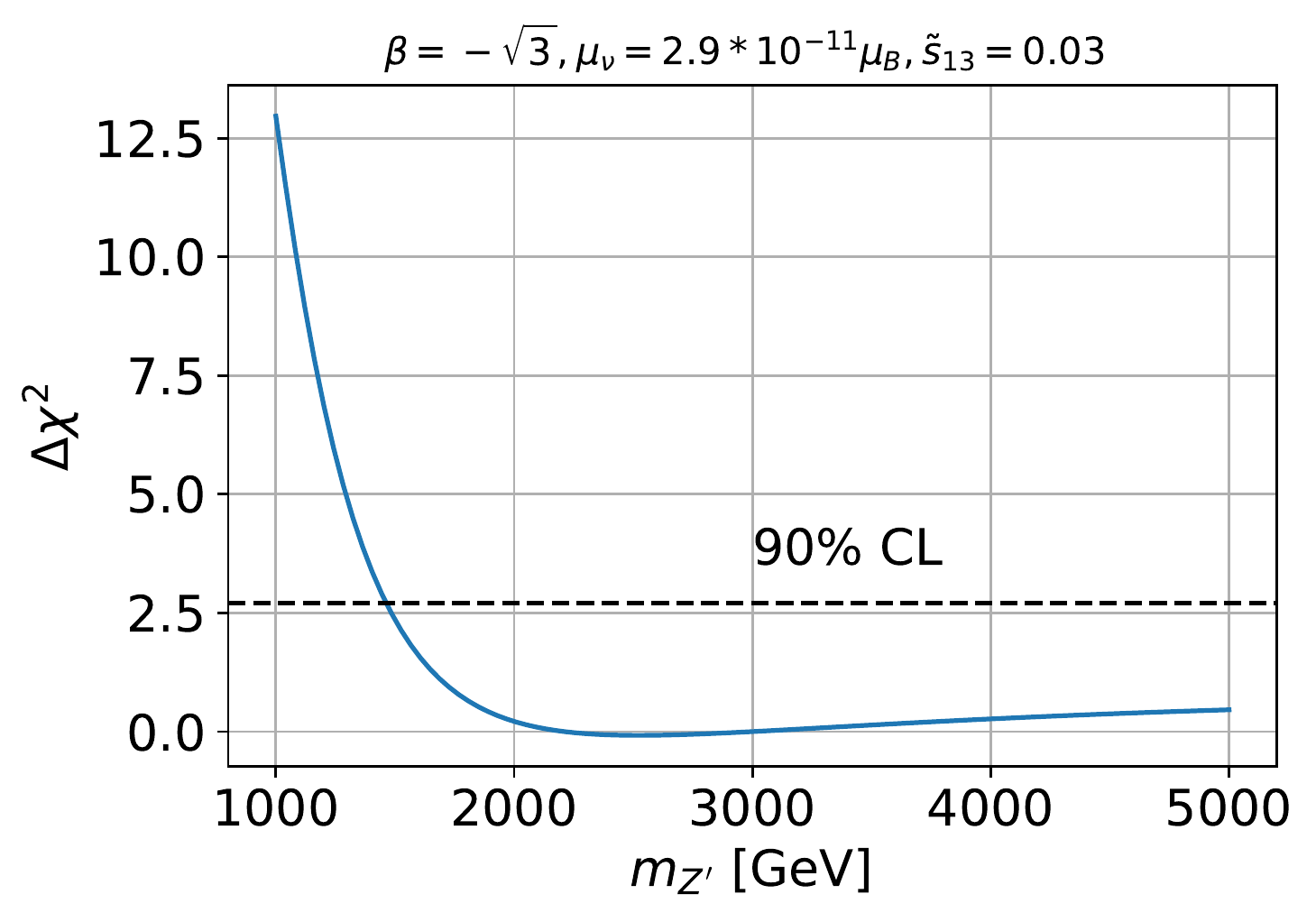}
\caption{$\Delta \chi^2$ profile of the sensitivity to the mass $m_{Z'}$}
\label{Deltachi2-mZp}
\end{figure}

\subsection*{Future CE$\nu$NS }
There are experiments on neutrino-nuclei scattering going on at COHERENT. In Table \ref{detector-subsystems} we summarize some of the detector subsystems at COHERENT. We will do the $\chi^2$ fit for these subsystems. 

\subsubsection*{Liquid Argon}

Recently, COHERENT collaboration report the first constraint on Coherent Elastic Neutrino-Nucleus Scattering in Argon. Two  analyses observed CEvNS event  over the background-only null hypothesis with greater than $3\sigma$ \cite{COHERENT:2020iec}. We will evaluate the $\Delta \chi^2$ fit for the data reported by liquid Argon detector.

The $\chi^2 $ is: 
\be 
\chi^2(\mathcal{P})= \min_{a} \Big[  \frac{(N_{exp}-N_{331\beta}[1+a])^2}{ \sigma_{stat}^2}+
\left( \frac{a}{\sigma_{a}} \right)^2  \Big],
\ee 
where $N_{exp}=159$ is the number of the measured events from the fit in Ref. \cite{COHERENT:2020iec}, $N_{331\beta}$ is the number of events predicted by $331 \beta$ model. The statistical uncertainty $\sigma_{stat}=\sqrt{N_{exp}+N_{BRN}}$ where $N_{ BRN} = 563$ represents the number of background events due to beam related neutrons (BRN).  The parameter a quantifies the normalization and $\sigma_a=8.5\%$ \cite{COHERENT:2020iec}.

\begin{figure}[ht!]
	\centering
	\includegraphics[scale=0.55]{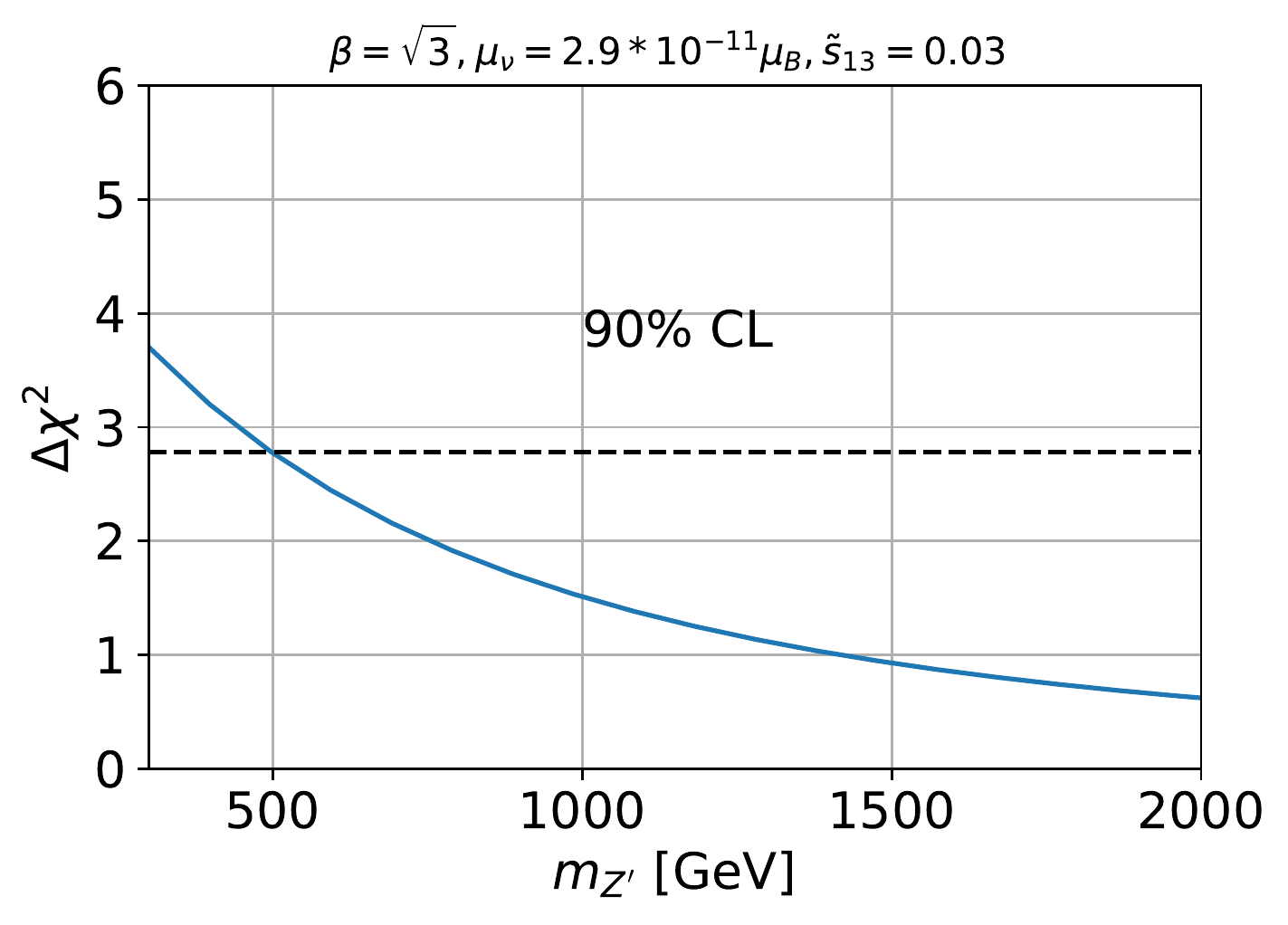}
	\includegraphics[scale=0.55]{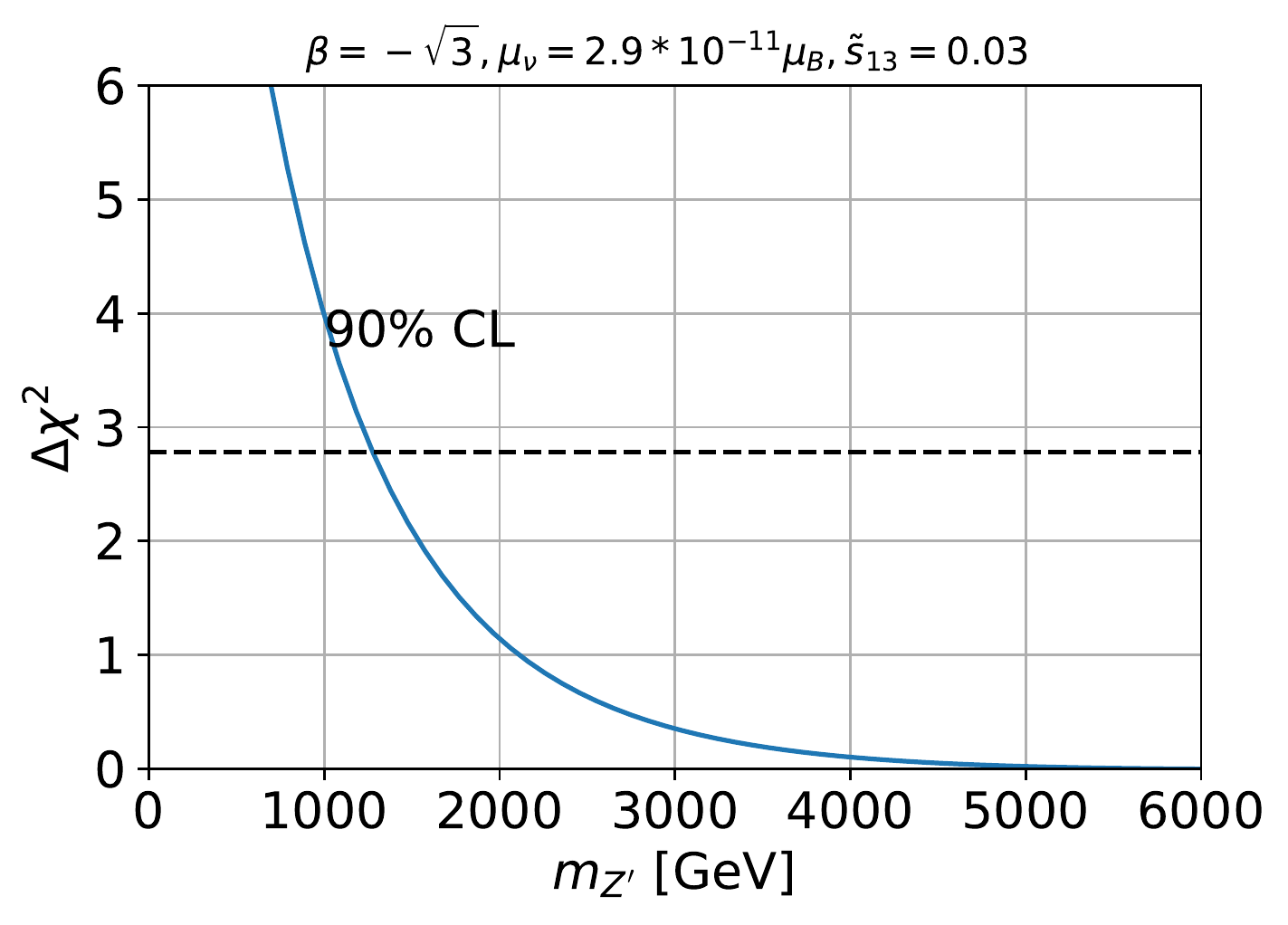}
	\caption{$\Delta \chi^2$ profile of the sensitivity to the mass $m_{Z'}$ for liquid Argon detector subsystem}
	\label{Deltachi2-mZp-ArEx}
\end{figure}

In Fig. \ref{Deltachi2-mZp-ArEx} we have evaluated the sensitivity  as function of the mass of the $Z'$ boson $m_{Z'}$ for the liquid Argon detector subsystem. The parameters of the detector are given as in Table \ref{detector-subsystems}. With the first data report by liquid Argon detector, we obtain $m_{Z'}\geq 0.5$ TeV for $\beta=\sqrt{3}$  at 90\% CL and   $m_{Z'}\geq 1.3$ TeV in the case $\beta=-\sqrt{3}$ at  90\% CL. These results are consistent and compliment previous contraints of CsI detector.

\subsubsection*{Germanium and NaI}

We consider a single nuisance parameter $\alpha$  for the systematic uncertainty  $\sigma_{sys} \in  [0.2,0.3 ] $.

The $\chi^2 $ in this case is 
\be 
\chi^2(\mathcal{P})= \min_{a} \Big[  \frac{(N_{SM}-N_{331\beta}[1+a])^2}{ \sigma_{stat}^2}+
\left( \frac{\alpha}{\sigma_{a}} \right)^2  \Big],
\ee 
where $N_{\mathrm{SM}}$ and $N_{331\beta}$ are the numbers of events predicted by the SM and  331$\beta$ model, respectively. The estimated statistical uncertainty is taken to be $\sigma_{stat} =\sqrt{N_{\mathrm{SM}} + N_{bg}}.$ The background is assumed to be flat and steady  $N_{bg} = \sigma_{bg} N_{\mathrm{SM}} $, with $\sigma_{bg}=0.2$. The $\chi^2$ fit is  evaluated for liquid Argon, Germanium and $NaI$ detector subsystem at COHERENT.

\begin{figure}[ht!]
\centering
\includegraphics[scale=0.55]{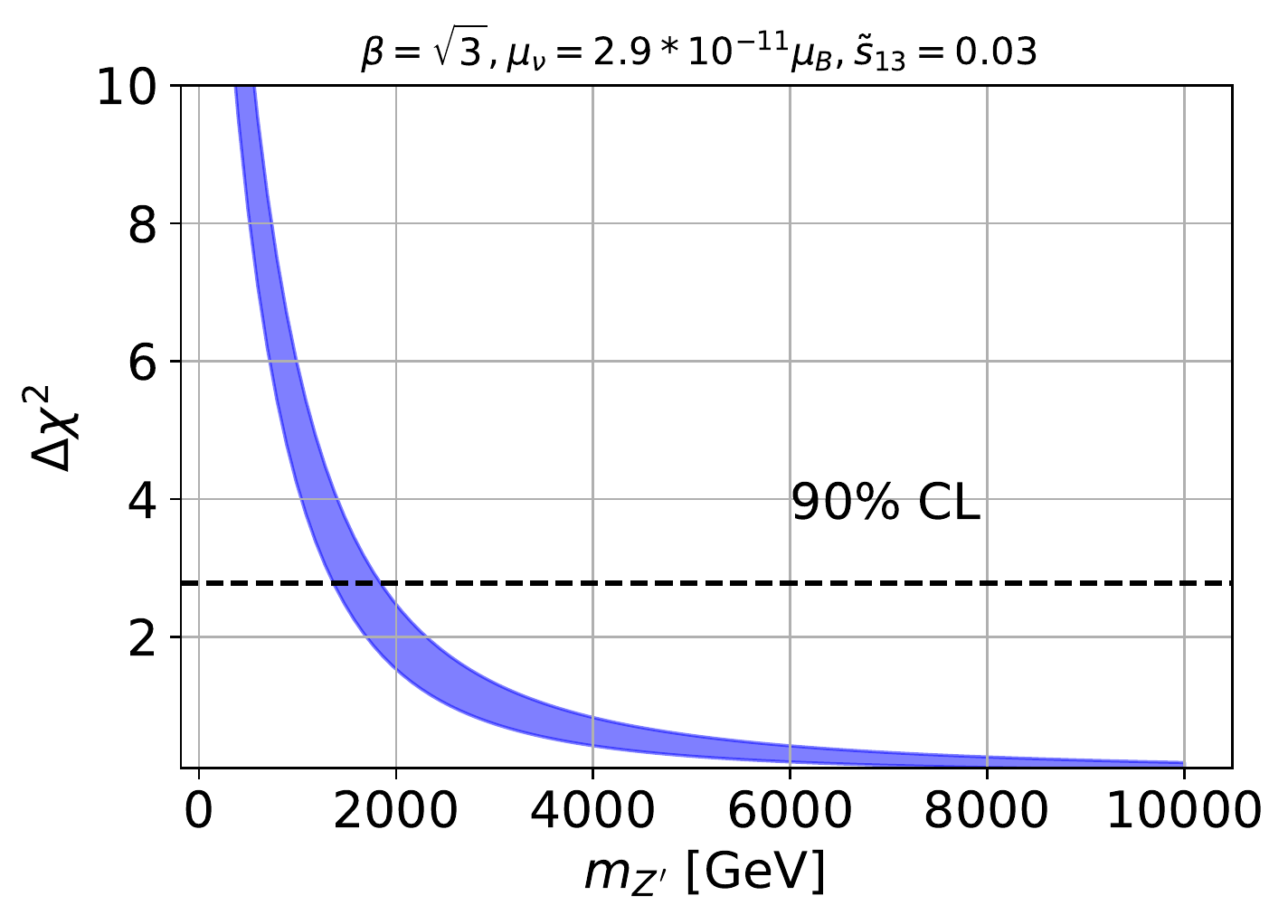}
\includegraphics[scale=0.55]{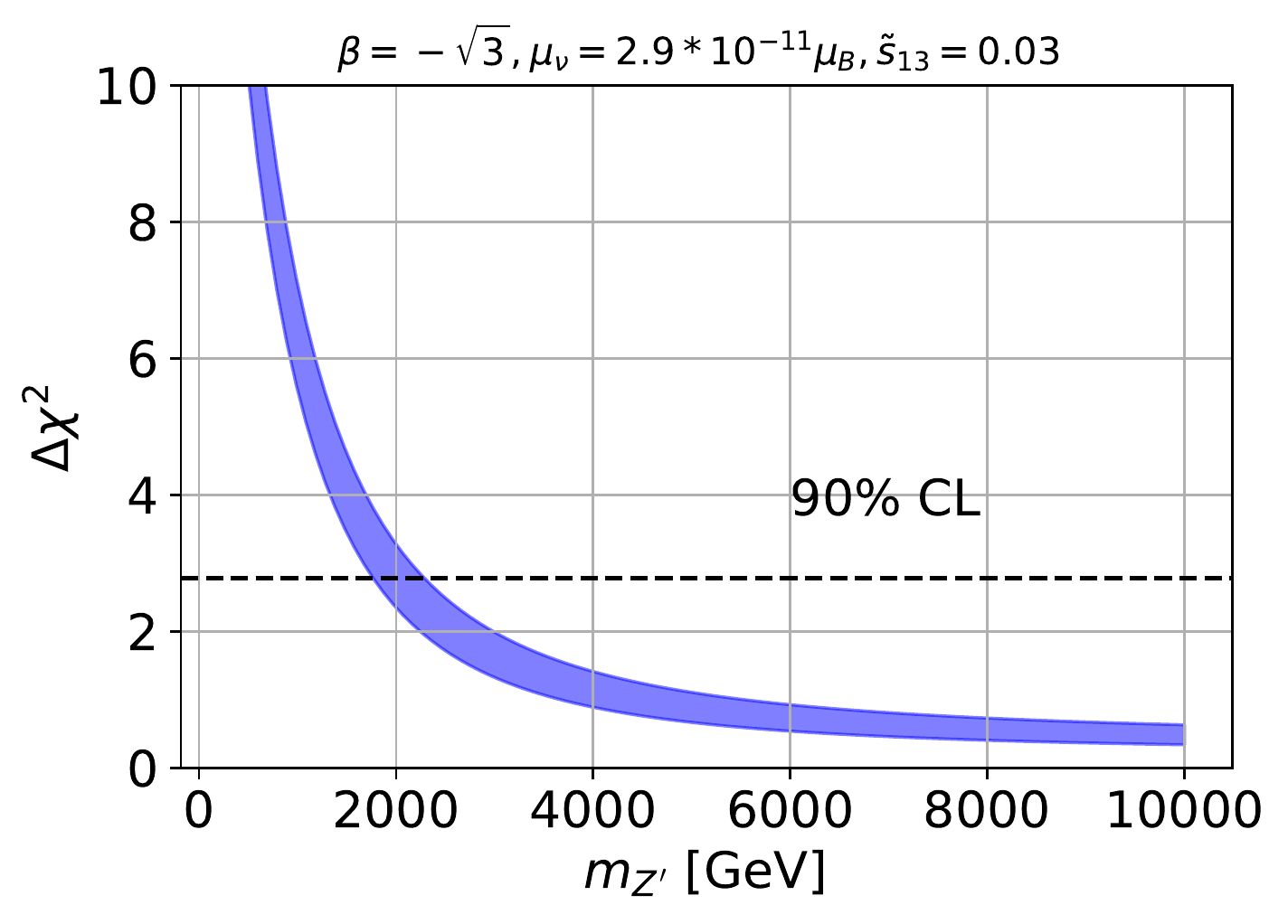}
\caption{$\Delta \chi^2$ profile of the sensitivity to the mass $m_{Z'}$ for Germanium detector subsystem}
\label{Deltachi2-mZp-GeEx}
\end{figure}

Next in Fig.\ref{Deltachi2-mZp-GeEx} we evaluate $\De \chi^2$  as a function of the mass of the $Z'$ boson $m_{Z'}$ for the Germanium detector subsystem. At 90\% CL $m_{Z'}\geq 1.9 $ TeV for $\beta=\sqrt{3}$ and   $m_{Z'}\geq 2.2$ TeV in the case $\beta=-\sqrt{3}$

\begin{figure}[ht!]
\centering
\includegraphics[scale=0.55]{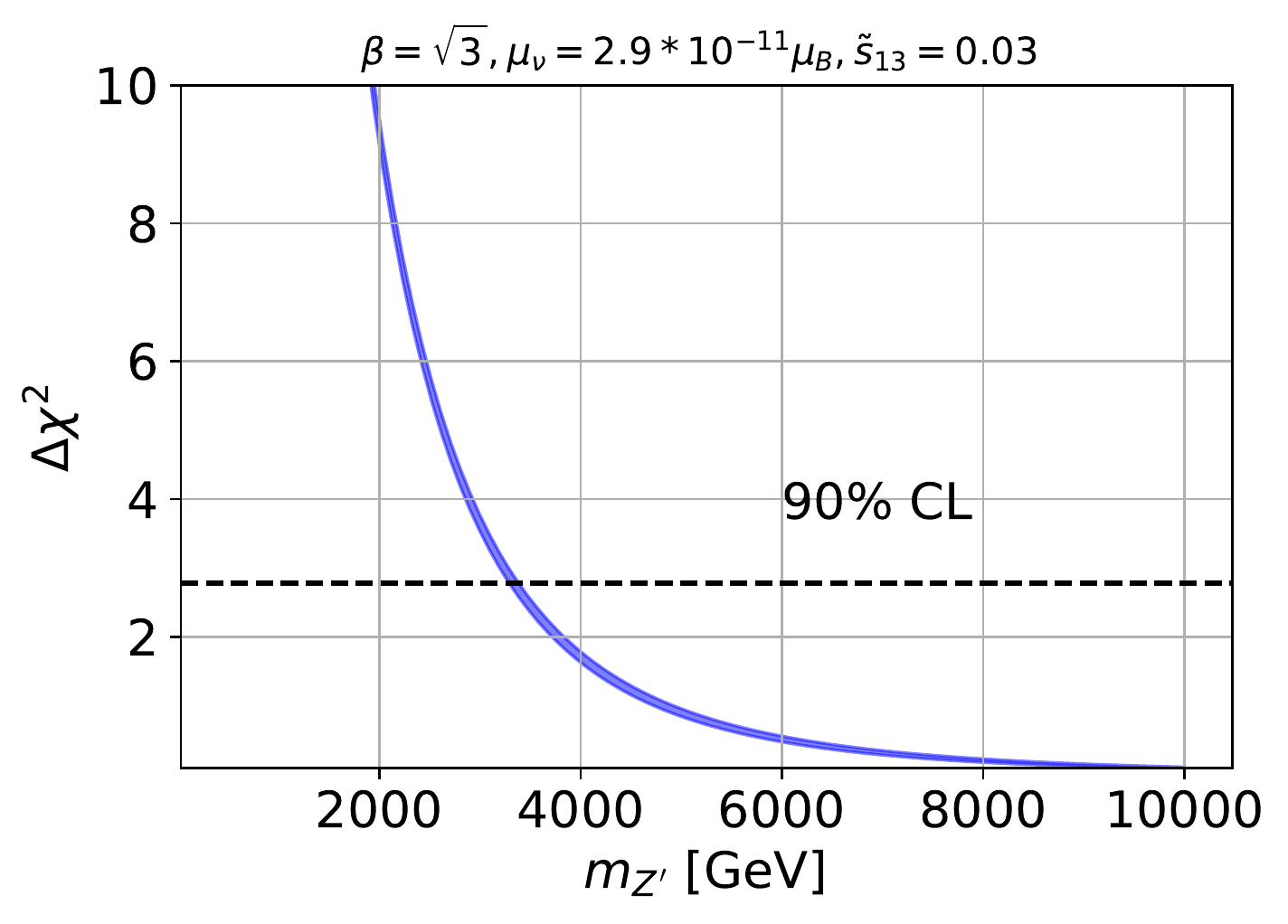}
\caption{$\Delta \chi^2$ profile of the sensitivity to the mass $m_{Z'}$ for NaI detector subsystem}
\label{Deltachi2-mZp-NaIEx}
\end{figure}

Finally in FIG. \ref{Deltachi2-mZp-NaIEx} we evaluated $\Delta \chi^2$ for NaI target detector. The mass of $Z'$ boson $m_{Z'}\geq  3.1 $TeV with 90\% C.L for $\beta=\sqrt{3}$. The projected sensitivity on the $Z'$ boson mass is improved compared with the sensitivity of $m_{Z'}$ given by CsI detector data. These results complement other bounds for the mass of Z' boson \cite{Nepomuceno2020,Profumo:2013sca,Long2019,Buras:2013dea,Buras:2014yna}.

\section{Conclusion}
\label{section5}

Studying the CE$\nu$NS process is an effective method to probe new physics effects at low energy.
In this work we have used the  experimental data of this process to discuss on the lower bound of $m_{Z'}$  predicted by the 331$\beta$ model. We have derived the effective Lagrangian of four fermion interactions of neutrinos and quarks, the  corresponding the weak charge in the  331$\beta$ model frameworks, then indicated that the corresponding  weak charge correction is large with large $|\beta|$.  Especially for largest allowed values of $\beta=\pm \sqrt{3}$, the deviation of the weak charges between the SM and $331\beta$ predictions is large at small $m_{Z'}$, which may leads to the inconsistency between the 331$\beta$ and experimental data. We showed that the expected number of events of the models is in agreement with data given by COHERENT experiment if $m_{Z'}$ is large enough. The sensitivities on  $m_{Z'}$   corresponding  two specific cases $\beta=\pm \sqrt{3}$ are evaluated. We found that the allowed values of the neutral gauge bosons mass are $m_{Z'}\geq 1.4$ TeV for $\beta=-\sqrt{3}$ with 90\% CL We perform the $\chi^2$ test for future CE$\nu$NS  liquid Argon, Germanium and NaI detector subsystems. Our analysis based on the first result reported by liquid Argon detector constraint $m_{Z'}\geq 0.5$ TeV for $\beta=\sqrt{3}$ and $m_{Z'}\geq 1.3$ TeV for $\beta=-\sqrt{3}$ with 90\% CL. For Germanium detector and NaI detector subsystems, the $\chi^2$ fit  indicated the favor range of the  $Z'$ boson mas   $ m_{Z'}\geq [2,3.1] $ TeV with 90\% CL.

Our results indicate that  low-energy
high-intensity measurements can provide a valuable probe, complementary to high energy collider searches at LHC and electroweak precision measurements.	



%
\bibliographystyle{utphys}
\bibliography{CEnuNS-331beta}

\end{document}